\documentstyle[12pt,epsf]{article}
\def\beq{\begin{equation}}   
\def\eeq{\end{equation}}
\def\lsim{\mathrel{\rlap{\lower3pt\hbox{\hskip0pt$\sim$}}
\raise1pt\hbox{$<$}}}         %less than or approx. symbol
\def\gsim{\mathrel{\rlap{\lower4pt\hbox{\hskip1pt$\sim$}}
\raise1pt\hbox{$>$}}}         %greater than or approx. symbol

\newcommand{\ra}{\rightarrow}

\newcommand{\Eq}[1]{Eq.\hspace*{.2em}(\ref{#1})}

\begin{document}

\begin{titlepage}
\renewcommand{\thefootnote}{\fnsymbol{footnote}}

\begin{flushright}
TPI-MINN-98/06-T\\
UMN-TH-1702/98\\
ICN-UNAM 98-04\\
May 1998 \\
hep-th/9806006 
\end{flushright}

\vspace{0.3cm}

\begin{center}
\baselineskip25pt

{\Large\bf 
Energy Reflection Symmetry of Lie-Algebraic Problems: Where the 
Quasiclassical and Weak Coupling Expansions Meet}

\end{center}

\vspace{0.3cm}

\begin{center}
\baselineskip12pt

{\large  M. Shifman}

\vspace{0.1cm}

Theoretical Physics Institute, University of Minnesota, Minneapolis,
MN 54555 

\vspace{0.2cm}

{\em and}

\vspace{0.2cm}

{\large A. Turbiner}

\vspace{0.1cm}
Instituto de Ciencias Nucleares, UNAM, 04510 Mexico D.F., Mexico

\vspace{0.1cm}

{\em and}

ITEP, Moscow 117259, Russia $^\dagger$

\vspace{0.7cm}

{\large\bf Abstract} \vspace*{.25cm}
\end{center}

We construct  a class of one-dimensional Lie-algebraic problems 
based on
$sl(2)$ where the spectrum in the algebraic sector has a dynamical 
symmetry $E\leftrightarrow - E$. All $2j+1$ eigenfunctions in the 
algebraic sector are paired, and inside each pair are related to each 
other by a 
simple analytic continuation $x\ra ix$, except the zero mode 
appearing if $j$ is integer. At $j\ra\infty$ the energy of the highest 
level in the algebraic sector can be calculated by virtue of the 
quasiclassical expansion, while the energy of the ground state can be 
calculated as a weak coupling expansion. The both series coincide 
identically.

\vspace{3cm}

\rule{2.4in}{.25mm} \\
$^\dagger$ On leave of absence.

\end{titlepage}

\newpage

{\em 1.  Introduction.}\hspace{0.7cm} A hidden  algebraic structure 
of the quasi-exactly solvable  (QES) Hamiltonians \cite{QES1} --
\cite{QES4} 
leads to  non-trivial dynamical properties
of the QES systems. One of such properties was observed in 
\cite{QESrev}: it was noted that all levels in the  algebraic sector of 
the simplest
QES
problem (see Eq. (\ref{1}) below) are
symmetric under 
$E\leftrightarrow - E$. This property  will be referred to as the {\em 
energy-reflection} 
(ER) symmetry. 
In the present paper  we derive  a class of one-dimensional QES 
Hamiltonian with this property and explore the consequences of the 
ER symmetry. A relation between the weak coupling and quasiclassical 
expansions will be established. 

One-dimensional QES problems are based on a (hidden) $sl(2)$  algebra;
they are characterized by  one quantized (cohomology) parameter $j$ 
where $j$ is 
half-integer  
\cite{ghko,Turbiner:1994}.
The 
number of levels in the algebraic sector is $2j+1$. 
In the systems to be constructed below 
 each state in  the algebraic sector 
with the energy eigenvalue $-E$ ($E>0$) is accompanied by a 
counterpart with   energy $E$,  if  $2j+1$ is even. If   $2j+1$ is odd,  a 
zero mode  exists 
while  
the remaining $2j$ levels come in pairs $\{ \psi_{-E}, \, \psi_{E}\}$. 
The eigenfunctions of the ER-symmetric  levels are 
related to each other by a straightforward analytic continuation 
$$
x\ra ix\, , \,\,\,  \psi_{E}\ra \psi_{-E}\, .
$$

At large $j$ the number of states in the algebraic sector is large. 
The highest levels  
still belonging to the algebraic 
sector can be regarded as highly excited states, and as such, are 
amenable to the quasiclassical treatment \cite{Bender}.
The parameter of the quasiclassical expansion is $1/j$.
At the same time, the lowest levels from  the algebraic sector are 
close to those of the harmonic oscillator. The anharmonicity is small,
and is  determined by a small parameter related to $1/j$.
Under the circumstances,  one can develop a standard weak-coupling 
perturbation theory and calculate $E$ as a series in the weak 
coupling.
Since the energy eigenvalues of the highly excited and low-lying  
ER-partners coincide, up to sign, the quasiclassical expansion 
and the weak coupling expansion in the QES problems 
with the ER symmetry {\em must be  identical}. 
We discuss how this identity is implemented, taking as a representative 
example
the ground state and its counterpart.

The simplest QES problem with the ER symmetry known for a 
long time \cite{QESrev}
is the 
sextic anharmonic oscillator, with a 
quantized coefficient in front of $x^2$, 
\beq
H= \frac{1}{2}\left[ p^2 + (x^6 - (8j+3)x^2)\right]\, , \,\,\, 
j=0,\frac{1}{2},1,
\frac{3}{2}, \, ... 
\label{1}
\eeq
In this case the algebraic sector consists of $2j+1$ levels of positive 
parity. 
The general QES potential possessing the 
ER symmetry involves certain elliptic functions and is related to some 
problems of practical importance. 

\vspace{0.2cm}

{\em 2. Generalities.} \hspace{0.7cm} The strategy we 
follow is described in \cite{QES1} (see also \cite{Turbiner:1994}) while 
the notation is borrowed from \cite{QESrev}.  
The generators of 
the $sl(2)$ algebra are defined as follows:
\beq
T^+ = 2j\xi -\xi^2\frac{d}{d\xi}\, , \,\,\,  T^0 = - j+\xi\frac{d}{d\xi}\, , 
\,\,\,  T^- = \frac{d}{d\xi}\, .
\label{2}
\eeq
If $j$ is a non-negative half-integer number,
a finite-dimensional irreducible representation exists,
\beq
R_{2j+1} = \{ \xi^0 , \xi^1 , ... , \xi^{2j}  \} \, ,
\label{3}
\eeq
where the subscript indicates the dimension of the representation. 
In general, the generators $T^{\pm}$ have the meaning of the raising 
(lowering) 
operators, 
$$
 R_{n} \stackrel{T^{\pm}}{\ra} R_{n\pm 1}\, , \,\,\, \mbox{while}\,\,\,  
R_{n} \stackrel{T^{0}}{\ra} R_{n}\, .
$$
The generic QES $sl(2)$-based Hamiltonian   is 
representable as a quadratic combination of the generators $T^\pm$ 
and $T^0$,
\beq
\hat{H} = \sum_{\pm ,0} \left( C_{ab}T^aT^b +C_a T^a \right) + C\, ,
\label{4}
\eeq
 where $C_{ab}, C_a, C$ are parameters.  One can always get rid of
$C_{0+}$ and $C_{-0}$ in favor of $C_{+0}$ and $C_{0-}$, respectively,  
due to the $sl(2)$ commutation relations. Moreover, 
 $C_{+-}$ and  $C_{-+}$ can be eliminated in favor of  $C_{00}$,  as a 
consequence of the irreducibility of the 
representation $R_{2j+1}$ (i.e. $T^+T^- + T^-T^+ +2T^0T^0 = 2j(j+1)$). The 
reference point for the energy is 
fixed by putting $C=0$.

After a change of variable and a (quasi)gauge transformation 
the operator $\hat{H}$ can be always reduced to the Schr\"{o}dinger 
form
\beq
\hat{H} \ra H\equiv e^{-a(\xi)} \hat{H} e^{a(\xi)}|_{\xi=\xi(x)} 
= -\frac{1}{2}\frac{d^2 }{dx^2}  + V(x)  \, .
\label{5}
\eeq
The key element in constructing the QES Hamiltonians with the 
ER symmetry is the following observation 
\cite{Turbiner1}: any tridiagonal matrix of the form
\beq
\left[ \begin{array}{cccccc}
0 & u_1 & 0 & 0 & ... & 0 \\
\ell_1 & 0 & u_2 & 0 & ... & 0 \\
0 & \ell_2 & 0 & u_3 & ... & 0 \\
... & ... & ... & ... & ... & ... \\
0 & 0 & 0 & ... & 0  & u_n \\
0 & 0 & 0 & ... & \ell_n  & 0 
\end{array}
\right]\, ,
\label{6}
\eeq
leads to the characteristic equation
\beq
E\, P_{n/2}\, (E^2) = 0\, , \,\,\, (n \,\,\, \mbox{even), and }\,\,\, 
\tilde{P}_{(n+1)/2}\, (E^2) = 0\, , \,\,\, (n \,\,\, \mbox{odd) }\, ,
\label{7}
\eeq
 where $P_{n/2}\, (z) $ and $\tilde{P}_{(n+1)/2}\, (z)$ 
are polynomials of $z$ of degree $n/2$ and $(n+1)/2$, respectively, and 
$n$ is defined in Eq. (\ref{6}). 
Thus, any matrix of the form (\ref{6}) guarantees the  
ER symmetry of the spectrum. It is evident, that the Lie-algebraic 
Hamiltonian (\ref{4}) has the matrix representation (\ref{6}) provided 
that the sum in \Eq{4} does {\em 
not} include the terms $T^+T^+$, $T^-T^-$, $T^0T^0$ and $T^0$. Thus, the 
most
general form of the Lie-algebraic Hamiltonian compatible with Eq. 
(\ref{6}) -- which, as was explained, ensures   the ER symmetry
in the algebraic sector -- is \footnote{The representation (\ref{6}) 
is a sufficient but not necessary condition. At some specific values of $n$ 
there may exist  QES systems with the ER symmetry which do not fall in 
the class of the systems we built.}
$$
\hat{H} = \alpha \, T^+T^0 + \beta \, T^0T^-  +\gamma \, T^+ + \delta \, 
T^- \, =
$$
\beq
A(\xi )\frac{d^2}{d\xi^2} + B(\xi )\frac{d}{d\xi } + C(\xi )\, , 
\label{9}
\eeq
where $\alpha , \beta , \gamma$ and $\delta$ are numerical 
constants, and $A,B,C$ are polynomials in $\xi$ of the third, second and 
first degree, respectively, 
\beq
A(\xi ) = - \alpha \xi^3 + \beta \xi\, , \,\,\,  B(\xi ) = [\alpha (3j-1)-
\gamma ] \xi^2 + (\delta - \beta j )\, , \,\,\,  C(\xi ) = 2 (\gamma j - 
\alpha j^2 )\xi \, . 
\label{10}
\eeq
Not all of the four  constants above represent physically interesting 
parameters. In general, two constants can be fixed by a
combination of  rescalings of the
variable $\xi$ and  the  energy. 
Using this freedom and starting from non-vanishing 
$\alpha$ and $\beta$ one can always reduce them to ``standard" 
$\alpha=\beta = 
-2$, see below. The parameters $\gamma$ and $\delta$ remain free.

Requiring the matrix (\ref{6}) to have 
 non-vanishing eigenvalues 
leads to a constraint that neither both parameters $(\alpha, \gamma)$ 
nor
both $(\beta,\delta)$ can be put to zero. One of the parameters in each 
pair can vanish, however. For instance, if $\alpha = 0$ the  general 
elliptic 
potential degenerates into a polynomial potential. Thus, the example 
presented in Eq. 
(\ref{1}) 
is nothing but a degenerate case of (\ref{9}) -- it corresponds to
$\alpha=0 , \,\,\,  \beta = \delta = -2 , \,\,\,  \gamma = - (2j+1) .$

Needless to say that $j$ is an additional  free parameter taking  a 
discrete set of values. Thus,  we deal with the 
three-parameter 
family of potentials: two 
continuous ones and one discrete. 

\vspace{0.2cm}

{\em 3. Elliptic potentials -- special case.} \hspace{0.7cm}
Prior to considering the general QES potentials with the ER 
symmetry we find it illuminating to  discuss a few 
representative examples. We start from
\beq
\alpha = \beta = - 2\,  , \,\,\, \gamma = - (8\nu + 6j +1)\, , \,\,\, 
\delta = - (2j+1)\, ,
\label{11}
\eeq
where $\nu $ is a constant. Since $\nu$ is free, so is $\gamma$; the 
parametrization of $\gamma$ above, in terms of $\nu$ and $j$, will be 
considered 
``standard".  Physical arguments (e.g. the stability of the potential) 
require $\nu$ to be 
non-negative. 
The parameter $\delta $ is fixed for the time being. Later on we will let 
$\delta$ vary too.

The physical variable $x$ in Eq. (\ref{5}) is determined by the inversion 
of the equation
\beq
\left( \frac{d \xi}{d x}\right)^2 = 4\xi - 4 \xi^3\, .
\label{dopone}
\eeq
(see \cite{QES1, QES3}). Equation (\ref{dopone}) has  solutions $-{\cal 
P}(x)$ and 
$1/{\cal P}(x)$ where ${\cal P}$ is the Weierstrass function. One could 
use either of them;  the second solution is more convenient for our 
purposes. Thus, 
\beq
\xi (x ) = \frac{1}{{\cal P}(x)}\, , \,\,\,  g_2 = 4\, , \,\,\, g_3 = 0\, ,
\label{12}
\eeq
where $g_{2,3}$ are the  
invariants of  the Weierstrass function. For the time being it is  assumed 
that $\xi \in [0,1]$ and 
$x\in
[0, x_*]$ where 
\beq
x_* = \int_0^1 \frac{d\xi }{2\sqrt{\xi - \xi^3}} = 
\frac{\sqrt{\pi}\Gamma (5/4)}{\Gamma (3/4)}\approx 1.311\, .
\label{13}
\eeq
Later on this constraint will be relaxed. Equation (\ref{12})
maps the interval $[0, x_*]$ onto $[0,1]$. The function $\xi (x )$
is double-periodic in the complex plane, with the periods $2x_*$ and 
$2ix_*$. Thus, under our choice of parameters,
the parallelogram of periods of the Weierstrass function becomes 
square.
The symmetry of the square immediately translates in the ER symmetry
of the quantal problem at hand. 
 We will use the fact that
\beq
\xi (- x ) = \xi (x )\, , \,\,\, \xi (i x ) = - \xi (x )
\label{14}
\eeq
stemming from the properties of the Weierstrass function with the 
above periods. The expansion of $\xi (x )$ at $x=0$ runs in powers of 
$x^2$. 

The phase $a(x)$ of the gauge transformation and the ``gauge 
potential" 
${\cal A}(x)$ are
\beq
a(x) = \left.
\frac{1}{4}\int \frac{(dA/d\xi )-2B}{A}\, d\xi\right|_{\xi = \xi(x)}= 
\left. 
-\nu \ln (1-\xi^2)\right|_{\xi = \xi(x)}\, ,
\label{15}
\eeq
and
\beq
{\cal A} = -\frac{1}{2}\frac{(dA/d\xi )-2B}{\sqrt{-2A}}=
\left. 
\frac{4\nu\xi^{3/2}}{\sqrt{1-\xi^2}}\, \right|_{\xi = \xi(x)}\, . 
\label{16}
\eeq
As a result, we get the  following  potential in  the 
Schr\"{o}dinger operator (\ref{5}):
\beq
V(x) = \left\{ \frac{4\nu (2\nu -1 )\xi^3}{1-\xi^2} -(8j^2 +2j+16\nu j+ 
6\nu )\xi \right\}_{\xi = \xi(x)}\, . 
\label{17}
\eeq
(The general formula for calculating the corresponding QES potential 
in the case at hand reduces to
$$
V(x) = \left\{ C(\xi ) + \frac{1}{2}({\cal A} (\xi ))^2 -\frac{d{\cal A} (\xi 
)}{d\xi}\sqrt{\xi - \xi^3 }
\right\}_{\xi = \xi(x)}\, ,
$$
where $C(\xi )$ is defined in Eq. (\ref{10}).)

This Schr\"{o}dinger problem is quasi-exactly solvable and can be 
considered beyond the original interval $[0, x_*]$. For $\nu = 0$
we deal with the periodic potential defined on the entire $x$ axis
(Fig. 1), which is akin to the Lam\'{e} problem \cite{Turbiner2}.
If $\nu >1/2$ the potential is singular at $x=\pm x_*$ (Fig. 2), and 
the problem is defined at $x\in (-x_*, x_*)$. The condition  $\nu 
>1/2$
is necessary for stability. 
The Hamiltonian changes sign under the transformation  $x\ra i x $, 
$$
H\ra - H\, ,\,\,\,  x\ra i x \, ,
$$
 as follows from Eq. (\ref{14}). The eigenfunctions $\{ \psi_E , \psi_{-
E}\}$ in  each 
pair interchange.  The ER 
symmetry is explicit. 
Let us  consider   separately two cases. 

\begin{figure}
\epsfxsize=8cm
\centerline{\epsfbox{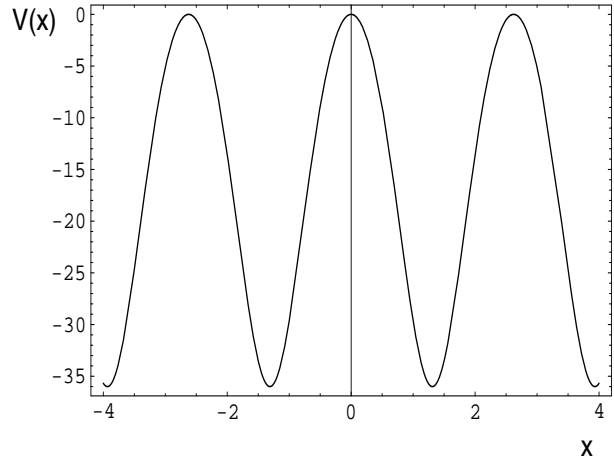}}
\caption{ The periodic QES potential of Eq. (\ref{17}) at $\nu =0$ and 
$j=2$.}    
\end{figure}

\begin{figure}
\epsfxsize=8cm
\centerline{\epsfbox{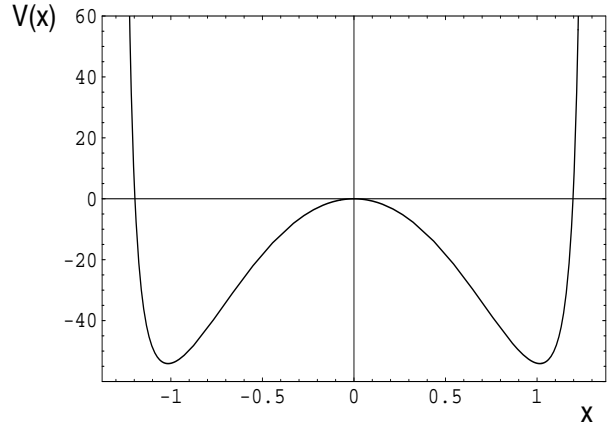}}
\caption{ The  QES potential of Eq. (\ref{17}) at $\nu =1$ and $j=2$
defined on the interval $-x_*<x<x_*$. The potential has a double-well 
shape. }    
\end{figure}

(i) $\nu = 0$.  The potential  and the eigenfunctions take the form 
\beq
V(x ) = -\frac{2j(4j+1)}{{\cal P}(x)} \, ,\,\,\,\,  \psi (x) = P_{2j} (\xi )\, 
.
\label{18}
\eeq
where $P_{2j}$ is a polynomial of degree $2j$.
At positive $E$ the spectrum is continuous, while
the counterpart wave functions $\psi_{-E}$
with negative energy eigenvalues  correspond to the boundaries of 
the Bloch zones; 
all these eigenfunctions are periodic. A very similar system, with a 
different coefficient in front of $1/{\cal P}(x)$, emerges at $\nu 
=1/2$.

(ii) $\nu > 1/2$.  The potential has a double-well form (Fig. 2). The 
singularity at 
$x\ra \pm x_*$ is of the form $(x\pm x_*)^{-2}$. 
The wave functions must vanish at $x=\pm x_*$, the spectrum is 
discrete. The algebraic sector includes the ground state and $2j$ 
excited states symmetric under $x\ra -x$. If $j$ is integer, one level 
lies exactly at zero, $j$ levels are below and above zero, respectively. 
If $j$ is half-integer, $(2j+1)/2$ levels lie below  zero and the same 
number above.

\vspace{0.2cm}

{\em 4. Generic elliptic potentials with the ER symmetry.} 
\hspace{0.7cm}
To proceed to the general case we invoke the only remaining freedom 
and let $\delta$ in Eq. (\ref{10}) float.    The following parametrization 
will be used:
\beq
\delta = - (2j+1) - \mu\, , 
\label{doptwo}
\eeq
while $\alpha ,\beta$ and $\gamma$ are the same as in  Eq. (\ref{11}).
If $\mu\neq 0$ Eqs. (\ref{15}) -- (\ref{17}) are modified as follows:
\beq
a(x) = \left\{
-\nu \ln (1-\xi^2)
-\frac{\mu}{8}\, \ln \, \frac{\xi^2}{1-\xi^2}\right\}_{\xi = \xi (x)}
\, ,
\label{dopthree}
\eeq
\beq
{\cal A} = \left\{
\frac{4\nu\xi^{3/2}}{\sqrt{1-\xi^2}}
-\frac{\mu \xi^{-1/2}}{2\sqrt{1-\xi^2 }}\right\}_{\xi = \xi (x)}
\, ,
\label{dopfour}
\eeq
and
$$
V(x) = \left\{ \frac{4\nu (2\nu -1 )\xi^3}{1-\xi^2} -(8j^2 +2j+16\nu j+ 
6\nu )\xi  +\right. 
$$
\beq
\left. 
\frac{\mu}{4\xi}\, \frac{1}{1-\xi^2}\, \left[ \frac{\mu}{2}-(1-3\xi^2)
\right]- \frac{2\nu\mu\xi}{1-\xi^2}\right\}_{\xi = \xi (x)}
\,  ,
\label{dopfive}
\eeq
where $\xi (x)$ is the same as before, see Eq. (\ref{12}). 
By inspecting this potential one concludes on physical grounds that
\beq
8\nu - 4 > \mu \geq 2\, .
\label{dopsix}
\eeq
Since the potential (\ref{dopfive}) is singular at $x\to 0$ (it explodes as 
$1/x^2$) the problem is defined for $\mu > 2$  on the interval $(0,x_*)$. 
If $\mu = 2$ there is no singularity at $x=0$, and the problem is defined
on the interval $(-x_*,x_*)$. The potential is depicted in Fig. 3. 

\begin{figure}
\epsfxsize=8cm
\centerline{\epsfbox{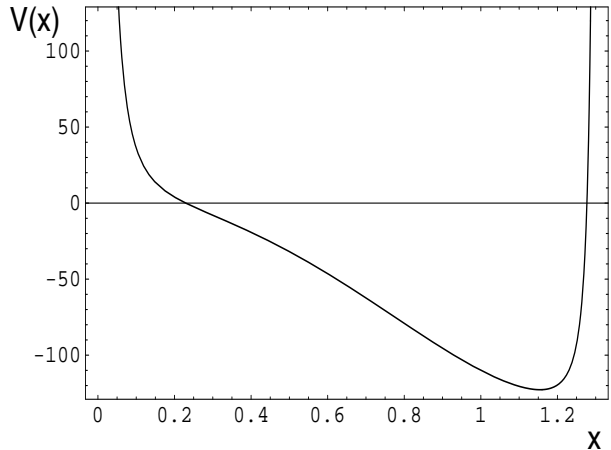}}
\caption{ The  QES potential of Eq. (\ref{dopfive}) at $\nu =1$, $\mu =3$  
and $j=3$
defined on the interval $0<x<x_*$. }    
\end{figure}

\vspace{0.2cm}

{\em 5}. $x\ra ix,\,\,\, H\ra - H$. \hspace{0.7cm}
Since the Hamiltonian we built possesses this property one might ask
why the ER symmetry is realized only in the algebraic 
sector 
rather than for the whole spectrum. From Fig. 2 it is quite obvious 
that
the   whole spectrum cannot have this symmetry -- the states 
at positive $E$ stretch  indefinitely, while for negative $E$
the lowest level is the ground state. Although the answer
to this question is rather obvious, an explanatory remark is in order. 

At arbitrary $E$ the second order differential equation $H\psi = 
E\psi$ has two linearly independent solutions, let us call them 
$\psi_{1,2}$. Generically both are non-normalizable. For $E=E_n$
one of the solutions (say, $\psi_1$) is normalizable, the other is not.
Generically, for arbitrary $n$,  the transformation $x\ra ix$
connects a normalizable solution at positive $E$ to a 
non-normalizable one at negative $E$. The latter does not lie in 
the physical Hilbert space, and there is no physical symmetry 
$E\leftrightarrow -E$. However, if $n$ belongs to the algebraic sector
the transformation $x\ra ix$ does not generate a non-normalizable 
solution, since the phase factor $a$ is invariant under $x\ra ix$.
Rather, in this case $x\ra ix$ connects a normalizable solution at 
positive $E$ to another normalizable solution at negative $E$.
Both belong to the physical Hilbert space, and $E\leftrightarrow -E$  
is a valid symmetry. 

The argument above is somewhat simplified, we cut corners. Although 
the conclusion is perfectly valid, the
careful treatment would require explicit introduction of the Stokes lines 
and consideration of sectors in the complex plane. It is important that in 
the problems with the ER symmetry under consideration,
one does {\em not}  jump from one branch onto another in the process 
of the analytic continuation from the purely real to purely imaginary 
values of $x$. In this respect the situation is different from that
discussed in \cite{AVT}, where the emphasis was on problems with the 
branch 
intertwining and the  leaps from one branch onto another. 

\vspace{0.2cm}

{\em 6. The zero mode.}  \hspace{0.7cm}
If $j$ is integer, there always exists a solution 
of the equation $\hat{H}\tilde\psi = E\tilde\psi$ with the vanishing 
energy eigenvalue, $E=0$. The corresponding wave function 
$\tilde{\psi}_0(\xi )$  contains only even powers of $\xi$, so that it is 
invariant under $x\ra ix$; it  can be given in the closed form,
\beq
\tilde{\psi}_0(\xi ) = \sum_{k=0}^j \frac{j!}{k!(j-k)!}\, 
\frac{[\gamma -\alpha j][\gamma -\alpha (j-2)] ... [\gamma -\alpha 
(j-2k+2)]}{[(j-1)\beta -\delta ][(j-3)\beta -\delta] ... [(j-2k+1)\beta 
-\delta]}\,\,  \xi^{2k}\, .
\label{zmcf}
\eeq

\vspace{0.2cm}

{\em 7. Quasiclassical vs.  weak coupling expansions. } \hspace{0.7cm}  
Consider the lowest and the highest levels in the algebraic 
sector in the limit $j\gg 1$. We will denote them by
$\psi_{-E_0}$ and $\psi_{E_0}$, respectively. The highest level is 
highly excited, and as such, is amenable to the quasiclassical 
treatment. 
The quasiclassical calculation of the energy of $\psi_{E_0}$
was carried out \footnote{Note that the unnumbered equation after 
Eq. (12) in \cite{Bender} contains errors in signs. These errors 
miraculously combine with another error -- the parameter of the 
quasiclassical quantization $n$ was taken to be $(\kappa +1)/2$ in Ref. 
\cite{Bender}, while actually it is $(\kappa -3)/2$ -- to annihilate each 
other.
The final expansion for the energy $E_0$ presented in Eq. (13) of Ref. 
\cite{Bender} is perfectly valid.  
Note that our $\kappa$ corresponds to $4J-1$ in \cite{Bender}. }
in Ref. \cite{Bender}. 
On the other hand, $\psi_{-E_0}$  corresponds to a system at the 
bottom of the well. This system is close to the harmonic oscillator, 
with a weak anharmonicity. One can develop the standard 
perturbation theory. The quasiclassical expansion and the weak 
coupling expansion have one and the same parameter, and coincide 
term by term, up to the overall sign \cite{QESrevprim}.

Although the assertion above is quite general and refers to all QES 
problems with the ER symmetry, we will elucidate it
using the simplest example. This will allow us to avoid bulky 
formulae.  The generalization to the general case is transparent.

As was noted in \cite{QESrev}, the simplest QES problem with the 
ER symmetry is that of Eq. (\ref{1}).  It is convenient 
to introduce a slightly different notation,
$$
\hat{H} = -2 T^0T^- - (2j+1)T^- - 2T^+\, ,
$$
\beq
H\psi \equiv \left\{ -\frac{1}{2}\, \frac{d^2}{dx^2} +\left[ 
\frac{x^6}{2} -
\frac{\kappa}{2}\, x^2 \right]\right\}\psi (x) = E\psi (x)\, ,
\label{sime}
\eeq
where
 $$
  \kappa = 8j+3 = 3,7,11,15, \, ... \, . 
$$
We are interested in the limit $  \kappa \ra\infty$. 

 At large $  \kappa$ the depth of the double-well potential becomes 
large
and the well width  small. 
The minima of $ V (x)$ lie at 
\beq
x = \pm x_0\, , \,\,\,\, x_0 
\equiv (\kappa /3)^{1/4}\, .
\eeq
Near, say, the right minimum
\beq
 V (x) =  -\frac{(\kappa )^{3/2}}{3\sqrt{3}} 
+ 2 \kappa  \left( x-x_0\right)^2 + 10 \left(\frac{\kappa}{3} 
\right)^{3/4} \, 
\left( x-x_0\right)^3 \, +...\, ,
\label{28}
\eeq
where the ellipses denote quartic and higher order terms; 
similar 
expansion is  
valid near the left minimum.  From Eq. (\ref{28}) 
it is easy to find the ground state energy in the form of an expansion 
in 
$1/\kappa $. Indeed, if we neglect exponential terms of the type
$\exp (- C\kappa )$ where $C$ is a positive constant, arising due to 
tunneling 
from one well into another \footnote{The impact of tunneling and the 
issue of the analytic structure in $\kappa$
are discussed in more detail in Sec. 8.}, the ground state level can be 
considered 
as that 
of the harmonic oscillator slightly perturbed by cubic, quartic, etc. 
terms.
The leading term in the ground state energy $E_0$ is just the 
classical energy
of the particle at rest  in the minimum, i.e. at $x_0$,
$$
-\frac{(\kappa )^{3/2}}{3\sqrt{3}} \, .
$$
The next-to-leading term is the zero-point oscillation energy of the 
harmonic 
oscillator,
$$
\frac{\omega}{2} = \kappa^{1/2}\, .
$$
Then  come the  corrections due to the 
anharmonic terms in the potential (\ref{28}). 
The first order correction due to the cubic term obviously vanishes.
Therefore, the next term in the $1/\kappa$ expansion of $E_0$ 
comes from  the quartic term in Eq. (\ref{28}) (treated as a first 
order 
perturbation), plus the second order 
perturbation generated by $10 \left({\kappa}/{3} \right)^{3/4}  
\left( x-x_0\right)^3$:
$$ 
\frac{135}{32}\, \frac{1}{3\sqrt{3}{\kappa}^{1/2}}\, -
\frac{275}{32}\, \frac{1}{3\sqrt{3}{\kappa}^{1/2}}, 
$$
respectively.  

Assembling all these  terms together we arrive at the following 
expansion for the ground state energy $-E_0$:
\beq
 -E_0 = -\left(\frac{\kappa}{3}\right)^{3/2}\left[ 1 - 
\frac{3\sqrt{3}}{\kappa
} +  \frac{35}{8\kappa^2} + O(\kappa^{-3})\right]\, .   
\label{33}
\eeq

So far only the lowest level  was discussed.
What can be said about the highest level in the algebraic sector?  

The ER symmetry  implies that the last 
level belonging to  the algebraic sector has the energy
\beq
E_0 = \left(\frac{\kappa}{3}\right)^{3/2}\left[ 1 - 
\frac{3\sqrt{3}}{\kappa
} +  \frac{35}{8\kappa^2} + O(\kappa^{-3})\right]\,  .
 \label{LJA}
\eeq
 Being considered as an 
excited state from the full set of states of the  Hamiltonian 
(\ref{sime}),  
this level should have been labeled by $(\kappa -3)/2$. Indeed, 
$\psi_{-E_0}$ is the ground state, then comes the first $P$-odd state,
the first $P$-even excitation, etc. The ground state and $2j$ $P$-even 
excitations belong to the algebraic sector. The last 
($P$-even) state from the algebraic sector has
\beq
n= \frac{\kappa - 3}{2}
\eeq
 where $n$ is  the number of zeros in the corresponding 
wave function. 

Let us discuss now how the very same expansion for $E_0$ emerges  
in the 
WKB approximation \cite{Bender}. 
It is instructive to  start from  the leading WKB 
approximation. Bohr and Sommerfeld's quantization rule at large $n$ 
implies
\beq
\int_b^a p dx = n\pi  = \frac{\kappa \pi }{2}\, , \,\,\,  \kappa  \ra\infty\, 
\label{BSQC}
\eeq
where 
\beq
p =\sqrt{2E -x^6 +\kappa x^2 }\, ,
\eeq
and $a$ and $b$ are the turning points.  We check that Eq. 
(\ref{BSQC}) is satisfied at $E=E_0 = (\kappa /3)^{3/2}$.
It is convenient to rescale the coordinate $x$,
\beq
          x = 2^{1/2} 3^{-1/4} \, \kappa^{1/4} y \, .
\eeq
Then 
\beq
p(E_0) = \sqrt{2}\, \left( \frac{\kappa}{3}\right)^{3/4}\,
\sqrt{1 - 4y^6 + 3 y^2} = \sqrt{2}\, \left( 
\frac{\kappa}{3}\right)^{3/4}\,
\sqrt{(1-y^2)(1+2y^2)^2} \, . 
\eeq
At $E=E_0$ the expression for $E-V$ factorizes, and the integral $\int 
pdx$ which in general is representable through elliptic functions
in fact reduces to elementary functions \cite{Bender}. 
Thanks to factorization we immediately see that the turning points
 are at $y=\pm 1$,
\beq
\int_b^a p dx  =\frac{4}{3}\, \kappa\, \int_0^1 dy\, (1+2y^2)\sqrt{1-
y^2} = \frac{\kappa \pi }{2}\, ,
\eeq
q.e.d. 

The first correction in the quasiclassical expansion can be calculated 
as easily as the leading term. 
Indeed, at this level the only change to be done is the substitution
\beq
n \ra n + \frac{1}{2}= \frac{\kappa}{2} - 1
\eeq
in the WKB quantization condition (\ref{BSQC}), and
\beq
E \ra E_0 = \left(\frac{\kappa}{3}\right)^{3/2}\left( 1-
\frac{C_1}{\kappa}
\right)\,  , 
\label{DEFC}
\eeq
where $C_1$ is a numerical coefficient, to be determined from the 
quantization condition
\beq
\int_b^a p(E_0)  dx = \left( \frac{\kappa}{2} - 1\right) \pi \, . 
\label{NQC}
\eeq
Now $p(E_0)$ takes the form 
$$
p(E_0) = \sqrt{2}\, \left( 
\frac{\kappa}{3}\right)^{3/4}\,
\sqrt{1 -\frac{C_1}{\kappa} - 4y^6 + 3 y^2} =
$$
\beq
\sqrt{2}\, \left( 
\frac{\kappa}{3}\right)^{3/4}\, \left[ \sqrt{(1-y^2)(1+2y^2)^2} -
\frac{C_1}{2\kappa}\, \frac{1}{\sqrt{(1-y^2)(1+2y^2)^2} }+ ...
\right]\, . 
\label{NEWP}
\eeq
We have already checked that the $O(\kappa )$ term in Eq. 
(\ref{NQC}) (it corresponds to keeping the first term 
in the 
square brackets) implies $E_0 = (\kappa /3)^{3/2}$. 
Matching of the $O(\kappa^0 )$ term in Eq. (\ref{NQC})
 (it corresponds  to the second  term in the square brackets) yields
\beq
C_1= 3\sqrt{3}\, ,
\eeq
in full accord with Eq. (\ref{LJA}).

Next-to-leading corrections in the quasiclassical expansion are 
calculated too \cite {Bender}, see also  footnote 1 above. 
The third and higher terms in the expansion require certain 
modifications of the WKB quantization condition which go beyond Eq. 
(\ref{NQC}).  From what we already know about the QES systems
under consideration, it is clear  that the $1/\kappa$ expansion of 
$E_0$ obtained through WKB must  match the weak 
coupling expansion. Six terms in the quasiclassical expansion of $E_0$ 
were found in Ref. \cite{Bender}.  Although it was expected, it was 
amusing to observe 
the coincidence with the first six terms in the weak coupling 
expansion. 

\vspace{0.2cm}

{\em 8. High-order behavior of the  expansion. } \hspace{0.7cm} 
It goes without saying that the weak coupling expansion (\ref{33})
is asymptotic. This is due to the possibility of the ``leakage" from the 
right to the left well. The high-order terms are factorially divergent and 
of the same sign. The behavior of the high-order terms in the
$1/\kappa$ series for the ground state energy is determined \cite{InAc} 
by the 
action 
of the instanton, the classical trajectory connecting the left 
and right minima in the Euclidean time. Let the instanton action  be
$S_0\kappa$, where $S_0$ is a number which we will calculate shortly. 
The Borel-resummed expression for the ground state energy
has the form
\beq
E_0 \sim \int  d g \frac{e^{-1/g}}{g - (2S_0\kappa )^{-1}}\, ,
\label{sone}
\eeq
where the principle value prescription  applies. The imaginary part of 
the integral is canceled by the imaginary part coming from the 
instanton-anti-instanton transition, which, in turn, is proportional 
\cite{InAc} to
$\exp (-2S_0\kappa )$. The condition of cancellation fixes the 
denominator of the integrand. Expanding Eq. (\ref{sone}) in $1/\kappa$ 
we
find the high-order tail of the $E_0$ expansion,
\beq
-E_0 \sim -\kappa^{3/2}\sum_{n>n_0} n! \frac{1}{(2S_0\kappa )^n}\, ,
\label{stwo}
\eeq
where $n_0$ is an integer large enough for the asymptotics to set in. 
The instanton action is readily calculable,
\beq
S_0\kappa =\int_{-x_0}^{x_0} dx \sqrt{\frac{2\kappa^{3/2}}{3\sqrt{3}}
+x^6 - \kappa x^2}\, ,\,\,\, x_0 =\left( \frac{\kappa}{3}\right)^{1/4}\, , 
\label{sthree}
\eeq
from where we obtain
\beq
S_0 =\ln \frac{1+\sqrt{3}}{\sqrt{2}}\approx 0.658 \, .
\label{sfour}
\eeq

The ER symmetry and Eq. (\ref{stwo}) imply that the very same 
factorial divergence is inherent to the quasiclassical expansion 
for energies of the highly excited states.  Certainly, this phenomenon is 
known in the literature \cite{EBB}. 
We find the 
argument above to be an illuminating way of demonstrating the 
asymptotic nature of the quasiclassical expansion. 
In fact, it is likely 
that the asymptotic regime starts quite early. Indeed, the first five 
coefficients in the quasiclassical expansion can be inferred from Ref. 
\cite{Bender}.  Denote  the coefficients 
 in front of ${1}/{(2S_0\kappa )^n}$ by $C_n$. Then, from  Eq. (13) of 
this work we get
$$
C_3 \approx 6.57\, , \,\,\, C_4\approx 20.2\, , \,\,\,  C_5\approx 117\, ,
$$
to be compared with the asymptotic prediction (\ref{stwo})
$$
C_3 =3!= 6\, , \,\,\, C_4=4!= 24\, , \,\,\,  C_5=5!= 120\, .
$$
Barring the possibility of a coincidental proximity, we conclude that
$n_0$ can be as low as three. 

The parameter $\kappa$ is related to the cohomology parameter and is 
quantized. 
The nature of the $1/\kappa$ expansions is closely related to the 
singularity structure in the complex $\kappa$ plane. 
In discussing this structure one should exercise caution, since 
the analytic continuation is performed from a discrete set of points,
$\kappa = 3,7, 11, ... $. This is one of the reasons why
the singularity structure in the complex $\kappa$ plane
turns out to be totally different  from that discussed in earlier works
\cite{AVT}, devoted to  the analytic continuation in continuous 
parameters in the
QES
problems. There are also some other reasons  responsible for the 
distinctions,  e.g. 
$\kappa$
appears as a coefficient of a subleading term in the potential, which is 
important.  
We do not dwell on this issue here, since it deserves a dedicated 
analysis. 
 
The quasiclassical quantization and the associated expansion imply  
$\kappa$ to be integer
(more exactly, 
$\kappa = 3,7, 11, ... $).  At the same time, the weak coupling expansion
(\ref{33}) is the same independently of whether or not $\kappa
\in \{3,7, 11, ...\}$. It holds for any sufficiently large $\kappa$.
Both expansions coincide order by order, to any finite order; yet if 
$\kappa\not\!\in
\{3,7, 11, ...\}$ the physical ER symmetry is absent, there is no reason 
for the coincidence of the absolute values of energy.
This means that the factorially divergent weak coupling series and the 
quasiclassical
expansion, presented in the square 
brackets in Eqs. (\ref{33}), (\ref{LJA}),  respectively,   define, generally 
speaking, 
two  distinct functions, despite the fact that the expansions {\em per se} 
are identical, order by order. The difference between these two 
functions is of 
the type
$\sin (\pi\kappa )\exp (-C\kappa )$; it vanishes at 
$\kappa = 3,7, 11, ... $. 
 For these and only these values of $\kappa$, making a full $2\pi$ 
circle in the complex plane around $\kappa =\infty$, starting from a 
positive $\kappa$ and returning to the very same point, we smoothly 
interpolate between  the lowest and  the highest levels in the algebraic 
sector; their positions  interchange. 
\vspace{0.2cm}

{\em 9.    The ER symmetry in the finite difference problems.} 
\hspace{0.7cm}
We have to mention that the energy reflection symmetry appears
also in  quantum-algebraic problems with  the Hamiltonians built from 
finite difference operators (such problems
naturally emerge in solid state physics). In order to display this
property let us consider, for instance, the  dilatation-invariant discrete 
operator $D_{\xi} $ defined as
\begin{equation}
\label{e3.1}
D_{\xi} f(\xi) = \frac{f(\xi) - f(q\xi)}{(1 - q) \xi}\, ,
\end{equation}
also known as  the  Jackson derivative (see e.g. \cite{e,gr}).
Here $q$ is a complex number. In the limit $q\to 1$  the  Jackson 
symbol obvioulsy goes into the conventional derivative.

Now, one can easily introduce \cite{ot}  a
finite-difference analog of the algebra of  the differential
operators (\ref{2}) based on the operator $D_\xi$,  instead of the 
continuous
derivative  (for a discussion see \cite{Turbiner:1994})
\begin{equation}
\label{e3.2}
\tilde  T^+ = \{n\}\xi-\xi^2 D_{\xi} \, , \qquad
\tilde  T^0 =  - \hat{n}+\xi D_{\xi}\, , \qquad
\tilde  T^- = \ D_{\xi} \, ,
\end{equation}
 where 
$$\{n\} = \frac{1 - q^{n}}{1 - q}$$ 
 is the so-called $q$ number, and 
$$\hat n \equiv \frac{\{n\}\{n+1\}}{\{2n+2\}}\, . $$

It is easy to check that  the operators (\ref{e3.2}) obey
the commutation relations of the quantum algebra $sl_{2q}$ for any
value of the parameter $j=n/2$ (see e.g.  \cite{z}). 
If $j$ is a non-negative
integer, the finite-dimensional representation (\ref{3}) of
the algebra (\ref{e3.2}) exists; it 
is irreducible when $q$ is not a prime root of unity. The same
line of reasoning  which we  followed to demonstrate the ER symmetry 
of the Hamiltonian (\ref{9}) can be  used in the case of the finite 
difference 
generators (\ref{e3.2}). In this way  we arrive at the conclusion that
the discrete Hamiltonian
$$
\hat{H} = \alpha \, \tilde T^+\tilde T^0 + \beta \, \tilde T^0\tilde T^-
 +\gamma \,\tilde T^+ + \delta \,\tilde T^- \, =
$$
\begin{equation}
\tilde A(\xi )D_{\xi}^2 + \tilde B(\xi )D_{\xi} + \tilde C(\xi )\, ,
\label{a.9}
\end{equation}
possesses the ER symmetry. Here $\alpha , \beta , \gamma$ and
$\delta$ are numerical constants, and $\tilde A,\tilde B,
\tilde C$ are polynomials of the third, second and first degree
in $\xi$, respectively,
\[
\tilde A(\xi ) = - \alpha q \xi^3 + \beta \xi\, , \,\,\,
\tilde B(\xi ) = [\alpha (\{n\}+\hat{n} -1)-\gamma ] \xi^2 +
(\delta - \beta \{n\} )\ ,
\]
\begin{equation}
\tilde C(\xi ) =  \{n\}(\gamma -
\alpha \hat{n} )\xi \ ,
\label{a.10}
\end{equation}
where $n=2j$.

It is remarkable that a particular form of this quantum-algebraic 
Hamiltonian (with a slightly different definition
of the discrete derivative)
appears in 
the Azbel-Hofstadter problem of the electron motion  on the 
two-dimensional lattice in the  transverse constant magnetic field
\cite{wz1,fn}. In this case the parameter $q$ is a prime root
of unity; it is related to  the magnetic flux through the lattice plaquette
(the flux is given by a rational number with an even denominator).

\vspace{0.2cm}

{\em 10. Comment on the literature. } \hspace{0.7cm} 
In Ref. \cite{KUW} a certain ``duality" transformation was 
suggested for the QES systems which inverts the signs of all levels 
belonging to the algebraic sector and, simultaneously, changes the form 
of the potential in a concerted way. It was observed that the potential 
(\ref{1}) is self-dual. Thus, the
 ER symmetry of the Schr\"{o}dinger problem  (\ref{1}) was 
rediscovered. 
It was noted then that the quasiclassical treatment of the QES problems 
should be qualitatively different from that of ``conventional" problems,
where there is no (quasi)exact solvability. The corresponding remark in 
\cite{KUW} is rather vague, and we feel that an explanatory 
remark is in order here. 

Suppose  the wave functions of a quantal system are treated in the WKB 
approximation. The WKB asymptotics, being considered in the complex 
$x$ plane,  
contains singularities at the points where the classical momentum 
vanishes. The Stokes lines are attached to these points; they divide the 
 complex $x$ plane into several sectors. The appropriate WKB expression 
for the wave function in the given sector, when analytically continued 
across a Stokes line, 
may or may not match  the appropriate WKB expression in another 
sector.  In other words, distinct  asymptotics may apply in the different 
sectors in the complex plane. This is a general situation. In the QES 
problems, for those levels that
are determined  algebraically,  the wave function is analytic everywhere 
except infinity
\footnote{It is implied that the original problem is defined 
on $x\in (-\infty , \infty )$, as in Eq. (\ref{1}). If the original problem is 
formulated on a finite interval, the wording must be changed 
appropriately.}. One and the same  asymptotics remains valid in all 
sectors; one can freely do analytic continuations across the Stokes lines.  
The singularities of separate  parts of the WKB expressions for the wave 
functions are superficial; 
they cancel when all parts are assembled together.
This property is well-known in the harmonic oscillator,  it extends to all 
QES systems, however.

The observation above belongs to A. Vainshtein. He pointed out
that the requirement of cancellation of these apparent singularities
can be used in order to generate  QES potentials. This requirement acts 
as a substitute of the algebraic structure within  the Lie-algebraic 
approach.

\vspace{0.2cm}

{\em Acknowledgments. }  \hspace{0.7cm} 
We are grateful to A. Vainshtein and M. Voloshin for 
useful and stimulating discussions. M.S.  
acknowledges an exchange of messages with Prof. M. Moshe.  A.T. would 
like to thank his colleagues from 
Theoretical Physics Institute, University of Minnesota, for warm 
hospitality.  

This work was supported in part by DOE under the grant number
DE-FG02-94ER40823.

\end{document}